\documentclass{sig-alt-release2}
\usepackage{paralist} \setdefaultitem{}{\textbullet }{ $\star$}{}
\usepackage{url}
\begin{document}

\title{Preliminary Analysis of Google+'s Privacy}
\subtitle{\small{(Poster)}}

\numberofauthors{2}

\author{
\alignauthor Shah Mahmood \\
        \affaddr{Department of Computer Science,}\\
        \affaddr{University College London,}\\
        \affaddr{United Kingdom}\\
        \email{shah.mahmood@cs.ucl.ac.uk}
 \alignauthor Yvo Desmedt \\
        \affaddr{Department of Computer Science,}\\
        \affaddr{University College London,}\\
        \affaddr{United Kingdom }\\
        \email{y.desmedt@cs.ucl.ac.uk}
 }

\conferenceinfo{CCS'11,} {October 17--21, 2011, Chicago, Illinois, USA.} 
\CopyrightYear{2011} 
\crdata{978-1-4503-0948-6/11/10} 
\clubpenalty=10000 
\widowpenalty = 10000

\pdfpagewidth=8.5in
\pdfpageheight=11in

\maketitle
\begin{abstract}
In this paper we provide a preliminary analysis of Google+
privacy. We identified that Google+ shares photo metadata with
users who can access the photograph and discuss its potential impact
on privacy. We also identified that Google+
encourages the provision of other names including maiden name, which
may help criminals performing identity theft. We show
that Facebook lists are a superset of Google+ circles, both
functionally and logically, even though Google+ provides a better user
interface. Finally we compare the
use of encryption and depth of privacy control in Google+
versus in Facebook.

\end{abstract}
\category{K.4.1}{Computer and Society}{Public Policy Issues - \textit{Privacy}}
\category{K.6.5}{Management of Computing and Information
  Systems}{Security and Privacy}
\terms{Security}
\keywords{Google+, Social Network, Privacy, Facebook}

\section{Introduction} \label{sec:Introduction}

Google launched its latest social networking
site Google+ on June 28$^{\rm th}$, 2011. According to comScore, an Internet traffic watcher, Google+
registered 25 million users in its first 5 weeks \cite{ComScore11},
which motivates a close scrutiny. Current
leader of social networking market and the key rival of Google+,
Facebook, has over 750 million registered users
\cite{FacebookStatistics11}. Facebook users share more than 30 billion
pieces of content (photos, videos, web links, notes, blog posts etc.) every month. 

Google+ like other social networks is used for sharing private
information including
status updates, occupation, employment history, home and work addresses, contact numbers, relationship status, photos, videos,
etc. As Google+'s market penetration grows, so will the
amount of data shared by its users. With the enormous amount of data produced on
social networks, privacy is one of the issues widely discussed both in
media and academia \cite{Anderson08}. Considering the importance of protection of the
private information of its users Google+ has introduced circles as a new concept to address the issue. 

Use of social networks has resulted in disclosure of
embarrassing information, loss of employment, suspension from school,
and blackmail \cite{Weiner11}. Social networks are
also used for social phishing attacks. Phishers harvest email
addresses to find the real names and social network profiles of their
victims \cite{Polakis10}. This harvest is possible because both
Google+ and Facebook require its users to use their real names and
allow search based on email addresses. Once
the real names and social network profiles are found, phishers extract
more information including people in the circles (or friend list) of
the victim, any comments, events attended etc. This information is
then used to craft personalized phishing attacks, called social
phishing \cite{Jagatic07}. Identity theft is costing US economy
\$15.6 billion a year
\cite{IdentityTheft11}. Moreover, social network status updates facilitated
robberies on several occasions, where the owner announced absence from
their property for a certain duration
\cite{FacebookRobbery10}. Furthermore, the large amount of data is also of
interest to advertisers and marketers. According to a survey by Social Media Examiner over 92\%
marketers use social networks as a tool
\cite{FacebookMarketer11}.

In view of the above discussion, it is very important and timely to analyze
Google+ and identify any privacy related issues. This is the
main goal of this paper. 

\textbf{Our contributions: }
\newline
\begin{compactitem}[\textbullet]
 \item We provide a preliminary analysis of privacy in Google+. We
   identify that Google+ shares the metadata of photos uploaded
   which could lead to privacy violations, discussed in Section
   \ref{PhotoMetadata}. Moreover, Google+ encourages its users to
   provide their past addresses and other names e.g. maiden name which
   could be used for identity theft. For further details see Section
   \ref{OtherNames}. 
 \item We compare Google+ circles (it's main privacy selling point) to
   Facebook lists. We show that, although Google+ circles have a
   better graphical user interface, they are logically and
   functionally a subset of Facebook lists. Details are provided in
   Section \ref{CirclesVSLists}.
\item We also make other comparisons between Facebook and Google+
  including the use of encryption and the ability to disable comments
  and message sharing. Further details are provided in Section \ref{OtherComparisons}
\end{compactitem} 

\section{Google+ Privacy}

In this section we present some privacy related problems and features of Google+.
We also make a comparison with Facebook, when applicable. 

\subsection{Google+'s photo metadata} \label{PhotoMetadata}
When a user uploads a photo on Google+, some metadata including the name of
the photo owner, the date and time the photo was taken, the make and model of
the camera etc. are made available to those with whom the photo is shared. This
set of information, in particular the date and time, may at first look
relatively innocent and trivial, but could in reality lead to some serious
privacy concerns. On August 10, 2007, in Pennsylvania (USA), a divorce lawyer
proved the spouse of the client being unfaithful to his partner, when the
electronic toll records showed him in New Jersey (USA) on that night and not
in a business meeting in Pennsylvania \cite{Divorce11}. With the metadata revealed by
Google+ a user might leak enough information to be legally held liable on
similar accounts.

Similarly, the make of the camera could be another concern for privacy. Higher
end cameras cost thousands of dollars. There have
been past incidents where the victims were killed for their cameras. In May 2011,
a Greek citizen, 44, was killed for his camera when taking his wife to the
hospital for child birth \cite{Camcorder11}.

\begin{figure} 
\centering
\epsfig{file=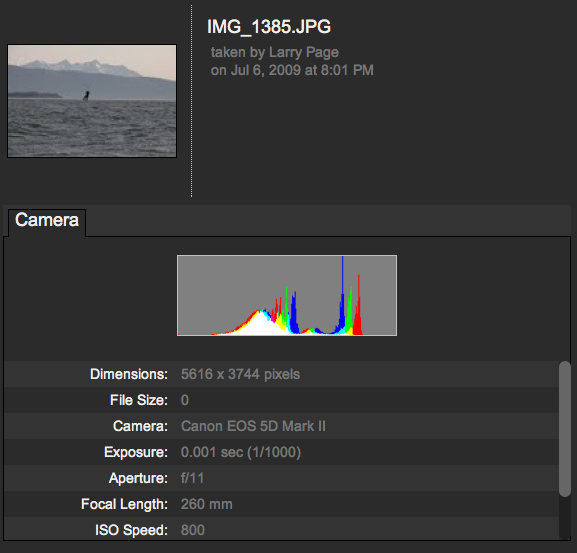, height = 3.1in, width = 3.3in} 
\caption{Metadata from a photo by Larry Page on Google+}\label{fig:Larry}
\end{figure}

Just to give an example of the level of information a picture exposes about
the camera, look at the metadata of the publicly shared pictures (from his Google+ profile) of Google co-founder
Larry Page, shown in Figure \ref{fig:Larry}. It reveals that
they he used a Canon EOS 5D Mark II camera to shoot his vacation
photographs. This camera is worth approximately USD 2000. This gives the robber
incentives.

\subsection{Cities lived in and other names on profile} \label{OtherNames}

In the ``About'' section of personal information, Google+ encourages
its user to 
provide the names of cities the user lived in and other names.
In the text box for other names, they write \textit{``For example: maiden name,
alternative spelling''}. Messages, photos and comments on social
networks and other online sources can be used to
infer family relationships. So, if someone can
link a profile to the profile of the mother and if the mother provides
the maiden name, then this could be used for identity theft, as
mother's maiden name is one of the most widely used secret question
\cite{Berghel00}. Moreover, the past addresses can only help the attacker with such
 attacks. 

\subsection{Google+ circles vs Facebook lists} \label{CirclesVSLists}

\begin{figure} 
\centering
\epsfig{file=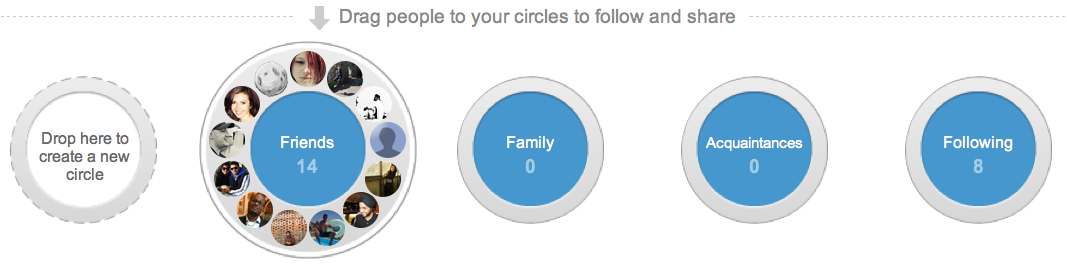, height = 1.1in, width = 3.3in} 
\caption{Google+ Circles} \label{fig:GoogleCircle}
\end{figure}

Paul Adams, then a Google employee, introduced the concept
of social circles \cite{Adams10}. These social circles act as the foundation
of circles in Google+. In Google+, by default there are four circles: ``friends'',
``family'', ``acquaintances'' and ``following''. We can remove/ rename any of the
default circles or add new circles. A user
can add any of her contacts to one or more circles just by a
simple drag and drop. Figure \ref{fig:GoogleCircle} shows the graphical
interface of Google+ circles. The intersection of two or more circles
can be a non-empty set. 

A user can share the content of her choice with a specific set of her
circles, all her circles, her extended circles(people in all her
circles and all people in the circles of the people in her circles)
and with the public (everyone). Google+ does not allow any exceptions,
i.e.\ , if some content is shared with a larger circle, there is no way
to exclude any subset of that circle. Anything shared with the public is
shared with all circles including the family and friends circle, which
might not be what the user may require. 

\begin{figure} 
\centering
\epsfig{file=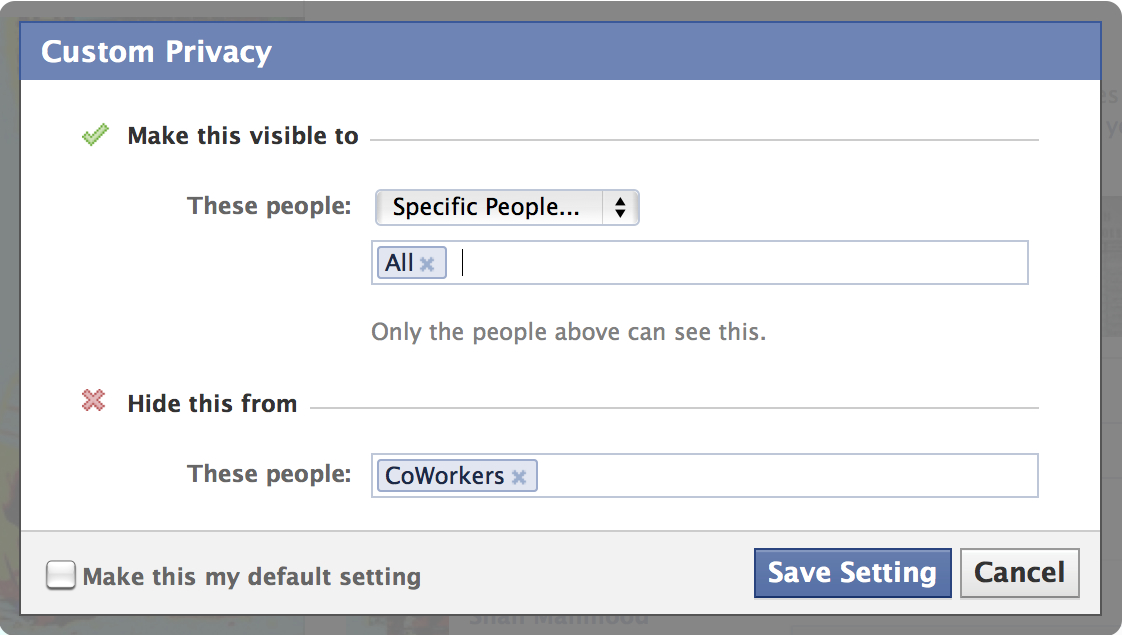, height = 1.7in, width = 3.3in} 
\caption{Share content with list``All'' but hide from list
 ``CoWorkers'' on Facebook} \label{fig:FBExceptions}
\end{figure}

Facebook on the other hand calls all the user's connections as
``friends''. Friends could be divided into groups
called ``lists''. There is no default list, so any structure has to be
created from scratch. Content on Facebook
can be shared with one or more lists, exactly like Google+ circles. But,
there is one difference that makes Facebook lists more robust than
Google+ circles i.e.\ the possibility of making exceptions. In
Facebook, we can limit access of our content to a list which is a subsets
of a set of lists with whom the content is shared. This means, we can
share a message with a list called ``All'' (containing all our
contacts) and still make the content invisibile to our ``CoWorkers'', as shown in Figure
\ref{fig:FBExceptions}. 

As Facebook's list creation was relatively cumbersome, recently a Facebook
application called ``Circle Hack'' \cite{CircleHack11} has been launched
which provides the Google+ circles graphical interface for Facebook
lists. The possibility and use of this application further proves our claim that Facebook lists are
logically and functionally a superset of Google+ circles.

\subsection{Google+ vs Facebook: other comparisons} \label{OtherComparisons}

Facebook uses an encrypted channel only for user authentication
(login) while Google+ uses it throughout the connection. This makes it harder to launch a
man in the middle attack against Google+. Moreover, Google+ allows
finer control of the content shared by a user. A user can disable comments
on a post at any time and enable it again later. This could be a
useful option to calm down any heated discussions, on the users wall,
between two contacts over the shared content or anything
else. Facebook, on the other hand, provides its users only with coarser control i.e.\ they
can only block a user from the entire wall but not on an individual content
basis (if it was initially shared with them). Furthermore, Google+ allows disabling
the resharing of a content at any instant on a content by content basis,
again its not possible in Facebook. Finally, Google+ allows its users to
edit their comments whenever they want. The time stamp of the last
editing remains visible on a comment, so users may modify or
backtrack their comments at any time. This too is not possible in
Facebook. 

\section{Related work}

Bradshaw identified the first privacy flaw in Google+ \cite{Bradshaw11}. The flaw was that any
content shared with a particular circle could be reshared with anyone
by someone from those circles. Although resharing of information is
always possible in the electronic world, if someone downloads a copy
and upload it again. But,
the simplicity and provision of a share button without proper
authorization is a privacy problem. This problem is now fixed by Google+. 

Social networks privacy and its potential threats have been widely
studied in recent years. One of the earliest works on potential threats to
individual's privacy including stalking, embarrassment and identity
theft was done by Gross \textit{et al.} \cite{Gross05}. 

Felt \cite{Felt07} presented a vulnerability in Facebook Markup Language
which lead to session hijacking. Bonneau and Dhingra independently
presented conditional and limited unauthorized access to Facebook
photos \cite{Bonneau09b,Dhingra08}.

\section{Conclusion} \label{Conclusion}
To conclude, we provided a preliminary analysis of Google+
privacy. We expressed concern that Google+ shares the metadata of the photos
uploaded by its users. We also showed that Google+ encourages its
users to provide their other names, e.g.\ , maiden names which may help
in identity theft. Moreover, we provided a comparison of Google+
circles with Facebook lists and showed that the latter is a superset of
the former, both logically and functionally even though Google+ provides
a better graphical interface. Finally, we provided other comparisons,
including the use of encryption and the possibility of modifying
comments at a later stage,
between Facebook and Google+.

\bibliographystyle{abbrv}
\bibliography{References_GooglePlus}



\end{document}